\documentclass[12pt,english]{article}
\pdfoutput=1
\usepackage{graphicx,array}
\usepackage{cite}
\usepackage{color}
\usepackage{latexsym}
\usepackage{amsthm}
\usepackage{amsmath}
\usepackage[titletoc]{appendix}
\usepackage{enumitem}
\usepackage{amssymb}
\usepackage[unicode=true,
 bookmarks=true,bookmarksnumbered=false,bookmarksopen=false,
 breaklinks=false,pdfborder={0 0 1},backref=false,colorlinks=true]
 {hyperref}
\hypersetup{pdftitle={A Nonlocal Approach to the Cosmological Constant Problem},
 pdfauthor={Sean M. Carroll and Grant N. Remmen},
 citecolor=blue,linkcolor=blue,urlcolor=blue,citecolor=blue,linkcolor=blue}
\usepackage{breakurl}
\usepackage[hang,flushmargin]{footmisc} 

\def\simgt{\mathrel{\lower2.5pt\vbox{\lineskip=0pt\baselineskip=0pt
           \hbox{$>$}\hbox{$\sim$}}}}
\def\simlt{\mathrel{\lower2.5pt\vbox{\lineskip=0pt\baselineskip=0pt
           \hbox{$<$}\hbox{$\sim$}}}}

\newcommand{\be}{\begin{equation}}
\newcommand{\ee}{\end{equation}}
\newcommand{\bea}{\begin{eqnarray}}
\newcommand{\eea}{\end{eqnarray}}
\newcommand{\Eq}[1]{Eq.~(\ref{#1})}

\newcommand{\Sec}[1]{Sec.~\ref{#1}}
\newcommand{\Secs}[2]{Secs.~\ref{#1} and \ref{#2}}

\newcommand{\Ref}[1]{Ref.~\cite{#1}}

\newcommand*\oline[1]{%
  \vbox{%
    \hrule height 0.5pt
    \kern0.68ex
    \hbox{%
      \kern-0.1em
      \ifmmode#1\else\ensuremath{#1}\fi
      \kern-0.1em
    }
  }
}

\definecolor{nicered}{rgb}{0.7,0.1,0.1}
\definecolor{nicegreen}{rgb}{0.1,0.5,0.1}
\hypersetup{
 colorlinks,citecolor=black,linkcolor=black,urlcolor=black}

\begin{document}
\baselineskip=14pt
\hfill CALT-TH-2017-016
\hfill

\vspace{2cm}
\thispagestyle{empty}
\begin{center}
{\LARGE\bf
A Nonlocal Approach to the \\Cosmological Constant Problem
}\\
\bigskip\vspace{1cm}{
{\large Sean M.\ Carroll and Grant N.\ Remmen}
} \\[7mm]
 {\it Walter Burke Institute for Theoretical Physics,\\
    California Institute of Technology, Pasadena, CA 91125}\let\thefootnote\relax\footnote{e-mail: \url{seancarroll@gmail.com, gremmen@theory.caltech.edu}} \\
 \end{center}
\bigskip
\centerline{\large\bf Abstract}

\begin{quote} \small
We construct a model in which the cosmological constant is canceled from the gravitational equations of motion. Our model relies on two key ingredients: a nonlocal constraint on the action, which forces the spacetime average of the Lagrangian density to vanish, and a dynamical way for this condition to be satisfied classically with arbitrary matter content. We implement the former condition with a spatially-constant Lagrange multiplier associated with the volume form and the latter by including a free four-form gauge field strength in the action.
These two features are enough to remove the cosmological constant from the Einstein equation. The model is consistent with all cosmological and experimental bounds on modification of gravity and allows for both cosmic inflation and the present epoch of acceleration. 
\end{quote}
	
\setcounter{footnote}{0}

\newpage
\tableofcontents
\newpage

\section{Introduction}\label{sec:Introduction}

The cosmological constant problem is the task of explaining the large hierarchy between the low observed value of the energy density of empty space and the high Planck scale of quantum gravity. One way of thinking about this problem is that, in a quantum field theory with a cutoff scale $\mu$, the energy density of the vacuum $\rho_\Lambda = \Lambda/8\pi G$ gets a contribution of order $\mu^4$ in the absence of tuning or some symmetry (e.g., supersymmetry). Famously, the discrepancy between the value of the cosmological constant $\Lambda_{\rm obs}$ inferred from cosmic distance measurements \cite{Riess:1998cb,Perlmutter:1998np} and the Planck scale $M_{\rm Pl} = 1/\sqrt{8\pi G}$ is $\Lambda_{\rm obs}/M_{\rm Pl}^2 \sim 10^{-120}$. While the presence of supersymmetry can ameliorate this tuning somewhat, even low-scale supersymmetry at $\mathcal{O}(\mathrm{TeV})$ would still leave a hierarchy of $\sim 10^{-60}$.

Many attempts have been made to solve the cosmological constant problem (see Refs.~\cite{Weinberg:1988cp,Carroll:1991mt,Bousso:2007gp} for reviews). Approaches have included allowing the cosmological constant to dynamically relax to (nearly) zero \cite{Abbott:1984qf,Brown:1987dd,Brown:1988kg,Feng:2000if,ArmendarizPicon:2000dh} and anthropic arguments \cite{Weinberg:1987dv,Weinberg:1988cp,Turok:1998he,Savas} positing a large multiverse populated with different local values of the vacuum energy. 
Other notable treatments of the problem include universe multiplication \cite{Linde:1988ws}, dividing the action by the volume of the universe \cite{Tseytlin:1990hn} (see also \Ref{Gabadadze:2014rwa}), the fat graviton \cite{Sundrum:2003jq}, self-tuning brane-world models \cite{Carroll:2001zy}, and the higher-dimensional bulk Casimir effect \cite{Elizalde:2002dd,Elizalde:2006iu}. 
Any proposed solution to the cosmological constant problem that involves classical fields adjusting a bare vacuum energy to some lower value must face up to a no-go theorem due to Weinberg \cite{Weinberg:1988cp}, which implies that no local field equations including gravity can have Minkowski solutions for generic values of the parameters.

One idea that has received considerable attention involves the introduction of a three-form gauge field $A_{\mu\nu\rho}$ with four-form field strength $F_{\mu\nu\rho\sigma}$, which in four spacetime dimensions contains no propagating degrees of freedom and contributes to the total vacuum energy \cite{Duff:1980qv,Hawking:1984hk,Baum:1984mc,Duff:1989ah,Duncan:1989ug}. The inclusion of such a gauge field in the action effectively makes the cosmological constant a constant of integration, since the value of the momentum of the three-form is completely free. Moreover, the gauge field can induce membrane nucleation, in which the cosmological constant is reduced in a stepwise fashion \cite{Brown:1987dd,Brown:1988kg}; this mechanism suffers the drawback that the instanton tunneling rate is exponentially suppressed, with the result that by the time the cosmological constant has been neutralized, the universe is devoid of matter and radiation \cite{Bousso:2007gp}. 
Other approaches, such as unimodular gravity (which degravitates the cosmological constant) \cite{Einstein:1919gv,vanderBij:1981ym,Buchmuller:1988yn,Henneaux:1989zc,Ellis:2010uc} and treating gravity as an equation of state \cite{Jacobson:1995ab}, similarly make the cosmological constant a constant of integration. 

While making the value of the cosmological constant a free parameter arguably represents progress, it still leaves the open problem of explaining the large hierarchy---why should this free parameter take on such a small value? An additional ingredient is needed. 
One possible such ingredient comes from Euclidean quantum cosmology \cite{Hawking:1984hk,Duff:1989ah,Baum:1984mc,Coleman:1988tj}. There, one considers the Euclidean path integral $S_{\rm E}$ over all of spacetime, positing that the probability of a given field configuration goes as $e^{-S_{\rm E}}$, setting the wave function for the universe to the exponential of the Euclidean effective action. For universes dominated by a cosmological constant, the action goes as $-1/\Lambda$, so for $\Lambda > 0$ the Euclidean path integral is dominated by small $\Lambda$, provided one invents a dynamical mechanism whereby $\Lambda$ can achieve different values, rather than being a fixed parameter. In the approach of  Refs.~\cite{Hawking:1984hk,Duff:1989ah,Baum:1984mc}, a four-form field strength constitutes the requisite mechanism for allowing $\Lambda$ to vary. In the approach of \Ref{Coleman:1988tj}, the dynamics of wormholes connecting larger universes were argued to make all constants of nature sample a distribution of possible values \cite{Giddings:1988cx,Coleman:1988cy}, effectively making $\Lambda$ a free parameter. 
Unfortunately, attempts to put the Euclidean path integral for gravity on a firm footing seem to run into problems 
\cite{Giddings:1988wv,Fischler:1988ia,Polchinski:1988ua}.

Still other approaches to the cosmological constant problem have included explicitly introducing nonlocality, as in \Ref{ArkaniHamed:2002fu}, which suggested a long-distance nonlocal modification of the field equations, and in the ``sequester'' approach of Refs.~\cite{Kaloper:2013zca,Kaloper:2014dqa,Kaloper:2015jra,Kaloper:2016jsd}. In the sequester models, the contributions of matter fields to the cosmological constant are avoided by coupling the matter sector to gravity through a rescaled metric whose scaling acts as a Lagrange multiplier. This Lagrange multiplier self-couples through a global term in the action that is not integrated over spacetime. As shown in \Ref{Kaloper:2015jra}, such models can be made local by promoting $\Lambda$ and $G$ to local fields and recasting the global terms as integrals over spacetime with respect to a different volume form.

In this paper, we describe a model for the cancellation of the cosmological constant that is classical, but nonlocal.
While our approach is somewhat ad hoc and phenomenological, it may serve as a pointer toward a more comprehensive theory. 
We use an arbitrary matter Lagrangian with the addition of a four-form field. For the action, we assume that its spacetime average value vanishes, which can be attained using a Lagrange multiplier that is constant over spacetime. Interestingly, this model achieves the removal of the cosmological constant from the gravitational field equations, while still allowing for the present era of accelerated expansion and without running afoul of any tests of general relativity or the equivalence principle. We present the model and its dynamics in \Sec{sec:model}, discuss its implications in \Sec{sec:implications}, and conclude in \Sec{sec:conclusion}.

\section{Theory and Dynamics}\label{sec:model}

\subsection{The model}

Any four-manifold with a Lorentzian metric $g_{\mu\nu}$ comes equipped with a natural volume form given by the Levi-Civita tensor $\boldsymbol{\epsilon} = \sqrt{-g}\, \mathrm{d}^4x$, where $g=\det(g_{\mu\nu})$. 
In components, $\epsilon_{\mu\nu\rho\sigma}=\sqrt{-g}\hat{\epsilon}_{\mu\nu\rho\sigma}$, where 
$\hat{\epsilon}_{\mu\nu\rho\sigma}$ is the Levi-Civita symbol, a density of weight 1 with $\hat{\epsilon}_{0123}=+1$.
Since $\boldsymbol{\epsilon}$ is a top form, any other four-form $\boldsymbol{\mu}$ can be written as a scalar field $\eta(x)$ times $\boldsymbol{\epsilon}$ and in principle any such four-form can be used as a volume element.

Our model is based on the assumption that the correct physical volume form used for defining the action of the theory is not necessarily the Levi-Civita tensor, but some other four-form that is also covariantly constant with respect to the metric:
\be
  \nabla_\lambda\mu_{\alpha\beta\gamma\delta} = \nabla_\lambda(\eta\epsilon_{\alpha\beta\gamma\delta}) = 0. 
  \label{eq:mu}
\ee
It is immediate that this requirement is equivalent to the parameter $\eta$ being a constant, rather than a spacetime-dependent field:
\be
  \nabla_\lambda \eta = 0.
\ee
We will discuss possible motivation for this model later, but for now we simply take as input this modification of the standard rules of the action formulation of general relativity. The upshot of this assumption is that there is a Lagrange multiplier $\eta$ in front of the action of the universe,
\be
S = \eta \int \mathrm{d}^4 x \sqrt{-g} \mathcal{L},\label{eq:etaaction}
\ee
enforcing that the total action vanish, 
\be 
\int \mathrm{d}^4 x \sqrt{-g}\mathcal{L} = 0.\label{eq:condition}
\ee
Had we instead allowed $\eta$ to be a spacetime-dependent field multiplying the Lagrangian density in \Eq{eq:etaaction}, this would have been equivalent to removing the requirement that the volume form $\boldsymbol{\mu}$ be covariantly constant with respect to the metric. Doing so would have resulted in an equation of motion for $\eta$ that sets the Lagrangian density to zero, rather than its spacetime integral, and moreover would have introduced propagating degrees of freedom into the four-form terms in the action, allowing for $F_{\mu\nu\rho\sigma}$ to vary with spacetime and hence not act as a cosmological constant.

Let's see what effect this mechanism has on the dynamics.
Consider a general matter Lagrangian $\mathcal{L}_{\mathrm{m}}$ plus gravity, with an arbitrary cosmological constant $\Lambda$. As usual, the cosmological constant is defined such that the on-shell Minkowski vacuum value of the action describing the matter and gravitational fields, given by $\mathcal{L}_{\mathrm{m}}$ and the Einstein-Hilbert term but not $\Lambda$, vanishes. To the action of the matter and gravitational fields we add the action for a free three-form $\mathbf{A}$ with a boundary term and with the coefficients fixed by canonical normalization,
\be 
S_{F}=\frac{1}{2}\int_{\mathcal{M}}\mathbf{F}\wedge\star\mathbf{F}
-\int_{\partial\mathcal{M}}\mathbf{A}\wedge\star\mathbf{F}.\label{eq:fourform}
\ee
Here, $\star$ is the Hodge dual defined with respect to the four-form $\boldsymbol{\mu}$ of \Eq{eq:mu}, so an overall factor of $\eta$ is implicit.
Such $n$-form gauge fields are generic in string theory and a classic item in the toolbox of the theorist wishing to address the cosmological constant. For a universe without boundary, we could imagine regularizing the action by integrating it up to some specified boundary surface. As noted in \Ref{Duncan:1989ug}, \Eq{eq:fourform} is the appropriate form of the four-form action to use when we are considering labeling the vacua by the value of $\star\mathbf{dA}$ on the boundary, which will be appropriate given the equation of motion we eventually derive for $\mathbf{A}$. Here, $\mathbf{F}=\mathbf{dA}$ or, in components, $F_{\mu\nu\rho\sigma}
=\left(\mathrm{d}A\right)_{\mu\nu\rho\sigma}
=4\nabla_{[\mu}A_{\nu\rho\sigma]}$.

The full action for our model is thus\footnote{We could also have added a boundary term to the gravitational part of the action, the Gibbons-Hawking-York term, which depends on the extrinsic curvature of the boundary of the manifold; however, as this term does not qualitatively affect our results, we drop it.}
\be 
S=\eta \int\mathrm{d}^{4}x\sqrt{-g}\left[\frac{1}{16\pi G}\left(R-2\Lambda\right)+\mathcal{L}_{\mathrm{m}}-\frac{1}{48}F_{\mu\nu\rho\sigma}F^{\mu\nu\rho\sigma}+\frac{1}{6}\nabla_{\mu}\left(F^{\mu\nu\rho\sigma}A_{\nu\rho\sigma}\right)\right].\label{eq:action}
\ee
The term $\Lambda$ appearing in \Eq{eq:action} is the total low-energy cosmological constant in the action, other than the contribution from $\mathbf{F}$. That is, in vacuum, $\mathcal{L}_\mathrm{m}$ vanishes on-shell, since the contributions of matter fields to the cosmological constant have already by definition been absorbed into $\Lambda$.
It will be useful to define $\mathcal{L}_{F}=-\frac{1}{48}F_{\mu\nu\rho\sigma}F^{\mu\nu\rho\sigma}$
and 
$\mathcal{L}_{DJ}=\frac{1}{6}\nabla_{\mu}\left(F^{\mu\nu\rho\sigma}A_{\nu\rho\sigma}\right)$, after Duncan and Jensen \cite{Duncan:1989ug}.

The basic effect of the Lagrange multiplier $\eta$ will be to force $F_{\mu\nu\rho\sigma}$ to take on a value such that it cancels the cosmological constant $\Lambda$ in the final equations of motion. We now proceed to see how this happens in practice.

\subsection{Equations of motion}

We will now analyze the equations of motion one obtains from the model (\ref{eq:action}) and ascertain their implications for gravitation. Let us introduce a regularization procedure for the volume of the spacetime $\mathcal{M}$, so that $\int \mathrm{d}^4 x \sqrt{-g} = V$.  The equation of motion for $\eta$ in \Eq{eq:condition} can be rewritten as 
\be
\frac{1}{V}\int \mathrm{d}^4 x \sqrt{-g}\mathcal{L} = \langle \mathcal{L} \rangle = 0,\label{eq:reg}
\ee
where angle brackets denote spacetime averages of the on-shell value. We can then extend this formalism to consider solutions with an infinite universe by taking the $V\rightarrow \infty$ limit in \Eq{eq:reg}. For our action \eqref{eq:action}, the dynamics of $\eta$ thus require
\be 
\frac{1}{16\pi G}\left(\left\langle R\right\rangle -2\Lambda\right)+\left\langle \mathcal{L}_{\mathrm{m}}\right\rangle -\frac{1}{48}\left\langle F_{\mu\nu\rho\sigma}F^{\mu\nu\rho\sigma}\right\rangle +\frac{1}{6}\left\langle \nabla_{\mu}\left(F^{\mu\nu\rho\sigma}A_{\nu\rho\sigma}\right)\right\rangle =0.\label{eq:constraint}
\ee
The cosmological constant, by definition, equals its own spacetime average value and hence does not require angle brackets.

We could have relaxed the assumption of the appearance of $\eta$ in \Eq{eq:etaaction}. To obtain our results, we only need the constraint in \Eq{eq:constraint}, which can be obtained from the weaker assumption that the spacetime average value of the action vanishes. For example, we could allow $S$ to take on some finite value, as long as the spacetime volume of the universe is infinite, and still reproduce \Eq{eq:constraint}.

Having concluded the preliminaries, which established the condition in \Eq{eq:constraint}, we now turn to the problem of deriving the dynamics of this model. The equation of motion for $\mathbf{A}$ is
\be 
\nabla_{\mu}F^{\mu\nu\rho\sigma}=0.\label{eq:eomF}
\ee
At this point, we could consider dualizing $\mathbf{F}$ to a scalar $\theta$ via
\be 
F_{\mu\nu\rho\sigma}=\theta\,\epsilon_{\mu\nu\rho\sigma}.\label{eq:dual}
\ee
The equation of motion \eqref{eq:eomF} implies that  $\theta$ is a constant, $\nabla_\mu\theta = 0$.
The addition of $\mathcal{L}_{DJ}$ means that this constant is (for the moment) arbitrary and not determined by the boundary values of $\mathbf{A}$.
Upon dualizing and using the identity $\epsilon_{\mu\nu\rho\sigma}\epsilon^{\mu\nu\rho\sigma}=-24$, we have the on-shell Lagrangians after substituting in the equation of motion \eqref{eq:eomF}:
\be 
\begin{aligned}\mathcal{L}_{F} & =\frac{1}{2}\theta^{2},\\
\mathcal{L}_{DJ} & =\frac{1}{6}\left(\nabla_{\mu}F^{\mu\nu\rho\sigma}\right)A_{\nu\rho\sigma}+\frac{1}{24}F_{\mu\nu\rho\sigma}F^{\mu\nu\rho\sigma}=-\theta^{2}.\label{eq:onshellL}
\end{aligned}
\ee
Under the dualization \eqref{eq:dual}, the constraint equation \eqref{eq:constraint} becomes
\be
\frac{1}{16\pi G}\left(\left\langle R\right\rangle -2\Lambda\right)+\left\langle \mathcal{L}_{\mathrm{m}}\right\rangle -\frac{1}{2}\theta^{2}=0,\label{eq:constraint:dual}
\ee
where \Eq{eq:eomF} means that we can drop the angle brackets on $\theta$, as it is constant.

For now, we will keep the action for the four-form written in terms of the gauge field $F_{\mu\nu\rho\sigma}$ for the purposes of computing the Einstein equation, i.e., the equation of motion for $g_{\mu\nu}$:
\be
\begin{aligned}
\frac{1}{\sqrt{-g}}\frac{\delta(\sqrt{-g}\mathcal{L})}{\delta g^{\mu\nu}} & =\frac{1}{16\pi G}\left(R_{\mu\nu}-\frac{1}{2}R\, g_{\mu\nu}+\Lambda\, g_{\mu\nu}\right)-\frac{1}{2}T_{\mu\nu}\\
 & \qquad+\frac{1}{96}g_{\mu\nu}F_{\alpha\beta\gamma\delta}F^{\alpha\beta\gamma\delta}-\frac{1}{12}F_{\mu\alpha\beta\gamma}F_{\nu}^{\;\;\alpha\beta\gamma}\\
 & =0.\label{eq:eomg}
\end{aligned}
\ee
We used that
\be 
\frac{1}{\sqrt{-g}}\frac{\delta}{\delta g^{\mu\nu}}\left(\sqrt{-g}F_{\alpha\beta\gamma\delta}F^{\alpha\beta\gamma\delta}\right)=-\frac{1}{2}g_{\mu\nu}F_{\alpha\beta\gamma\delta}F^{\alpha\beta\gamma\delta}+4F_{\mu\alpha\beta\gamma}F_{\nu}^{\;\;\alpha\beta\gamma}
\ee
and we did not need to consider the total derivative term 
$\nabla_{\mu}\left(\sqrt{-g}F_{\mu\nu\rho\sigma}A^{\nu\rho\sigma}\right)$, as its variation automatically vanishes.
In \Eq{eq:eomg}, $T_{\mu\nu}=-2\left(-g\right)^{-1/2}\delta(\sqrt{-g}\mathcal{L}_{\mathrm{m}})/\delta g^{\mu\nu}$ as usual.
 
Let us now dualize the four-form in \Eq{eq:eomg} according to \Eq{eq:dual}.
Using the identity $\epsilon^{\alpha\beta\gamma\mu}\epsilon_{\alpha\beta\gamma\nu}=-6\delta_{\nu}^{\mu}$, we have $F_{\mu\alpha\beta\gamma}F_{\nu}^{\;\;\alpha\beta\gamma}=-6\theta^{2}g_{\mu\nu}$ and hence the Einstein equation becomes
\be 
\frac{1}{16\pi G}\left(R_{\mu\nu}-\frac{1}{2}R\, g_{\mu\nu}+\Lambda\, g_{\mu\nu}\right)-\frac{1}{2}T_{\mu\nu}+\frac{1}{4}g_{\mu\nu}\theta^{2}=0.\label{eq:eomg:dual}
\ee
Finally, we may substitute the spacetime average condition in \Eq{eq:constraint:dual} into the Einstein equation \eqref{eq:eomg:dual} to cancel the cosmological constant:
\be 
R_{\mu\nu}-\frac{1}{2}R\, g_{\mu\nu}+\frac{1}{2}\left\langle R\right\rangle g_{\mu\nu}=8\pi G\,\left(T_{\mu\nu}-\left\langle \mathcal{L}_{\mathrm{m}}\right\rangle g_{\mu\nu}\right).\label{eq:Einstein}
\ee
This is the important result of our model: a set of gravitational field equations in which the cosmological constant has been dynamically removed.

\subsection{Solutions in general relativity}\label{sec:GR}

Let us assess the effects of the modified Einstein equation \eqref{eq:Einstein}.
First, we note that all quantities in angle brackets are by definition constants, as they are averages over the entire spacetime.
As a result, conservation of  $T_{\mu\nu}$---i.e.,  $\nabla^{\mu}T_{\mu\nu}=0$---follows in \Eq{eq:Einstein} from the Bianchi identity $\nabla^{\mu}R_{\mu\nu}=\nabla_{\nu}R/2$ in the same way as in the usual Einstein equation.
Furthermore, because of the constancy of $\langle R\rangle $ and $\langle \mathcal{L}_{\mathrm{m}}\rangle$, they could not have any local gravitational effect and therefore are immune to, e.g., solar system tests of general relativity and do not induce any equivalence principle violation.
Effectively, $\langle R/2 + 8\pi G\,\mathcal{L}_{\mathrm{m}}\rangle$ acts as a cosmological constant, but one that would vanish for vacuum configurations of the matter field, since by \Eq{eq:Einstein}, we have 
\be
\left\langle \frac{1}{2}R+8\pi G\,\mathcal{L}_{\mathrm{m}}\right\rangle =8\pi G\left\langle \frac{1}{2}T-\mathcal{L}_{\mathrm{m}}\right\rangle,
\ee
where $T = g^{\mu\nu}T_{\mu\nu}$. Thus, Minkowski space is a consistent solution of \Eq{eq:Einstein}.
Moreover, since this is truly a cosmological constant, the Friedmann equations and other general relativity solutions are unmodified in \Eq{eq:Einstein}.

\subsection{Spacetime averages of the gravity and matter actions}\label{sec:averages}

Let us now consider what values $\langle R\rangle $ and $\langle \mathcal{L}_{\mathrm{m}}\rangle$ would realistically take.
 
By definition, $\mathcal{L}_{\mathrm{m}}$ vanishes on-shell in vacuum, since we explicitly pulled out the cosmological constant in the action \eqref{eq:action}.
Moreover, for a universe dominated by fermionic matter and radiation, one has the Lagrangian for quantum electrodynamics (and possibly the gluon kinetic terms, which we will ignore in this discussion).
For radiation, $\mathcal{L}_{\mathrm{m}}=-\frac{1}{4}F_{\mu\nu}F^{\mu\nu}$ and so vanishes on-shell for a thermal bath, in which $\mathcal{L}_{\mathrm{m}}=\frac{1}{2}(E^{2}-B^{2})=0$.
Further, for massive fermionic matter, $\mathcal{L}_{\mathrm{m}}=\bar{\Psi}(i\not\!\! D-m)\Psi$.
Varying the Lagrangian with respect to $\bar{\Psi}$, we obtain the Dirac equation $(i\not\!\! D-m)\Psi=0$; since this is linear in $\Psi$ it is straightforward to substitute back into $\mathcal{L}_{\mathrm{m}}$, from which we find that $\mathcal{L}_{\mathrm{m}}$ also vanishes on-shell for massive fermionic matter (e.g., baryonic matter).
The only situation relevant to cosmological tests of gravity in our universe where $\mathcal{L}_{\mathrm{m}} \neq 0$ is a background (non-radiation) electromagnetic field, which is a very subdominant contribution to the energy budget of the universe at the present epoch and would be negligible for a universe that keeps expanding for a long time.
Indeed, for matter content that is eventually diluted away by the expansion of the universe (i.e., anything other than a cosmological constant or some
 phantom-energy-like field), $\langle \mathcal{L}_{\mathrm{m}}\rangle$ would vanish if the universe keeps expanding forever. Further, if the on-shell value of $\mathcal{L}_{\mathrm{m}}$ is nonzero only in a measure-zero portion of the universe, e.g., in a finite region of a spatially-infinite universe, then $\langle \mathcal{L}_{\mathrm{m}}\rangle$ also vanishes.

Therefore ignoring $\langle \mathcal{L}_{\mathrm{m}}\rangle$, we turn to the question of $\langle R\rangle$, which for vanishing $\langle \mathcal{L}_{\mathrm{m}}\rangle$ satisfies $\langle R\rangle =8\pi G \langle T\rangle$.
For a perfect fluid, $T=-\rho+3p$.
As long as $w\leq\frac{1}{3}$ and $\rho\geq0$, at least averaged over all of spacetime, then $\langle T\rangle \leq 0$ and we effectively have a negative cosmological constant in \Eq{eq:Einstein}, which corresponds to an effectively negative vacuum energy density.
Of course, if $T$ is only nonzero over a finite spacetime region and if the universe's spacetime volume is infinite, then $\langle R\rangle$
vanishes and we simply have the Einstein equation with no cosmological constant, $R_{\mu\nu}-\frac{1}{2}R\, g_{\mu\nu}=8\pi G\, T_{\mu\nu}$.
This would not be in conflict with the observed acceleration of the universe, since the present acceleration could be driven by a quintessence field that will eventually (though at arbitrary late time) turn off, leaving a finite $\langle R\rangle$ despite exponentially expanding the universe for an arbitrarily long, but not infinite, time. In exactly the same way, a finite period of inflation is completely possible in our model and further would manifest no differences from inflation in standard general relativity.

Moreover, if the universe undergoes phase transitions in a given cosmology, the exponential expansion associated with the false vacuum still occurs in our model via \Eq{eq:Einstein}, since these temporary effects do not impact spacetime-averaged quantities.  The present weak apparent cosmological constant observed in our own universe then has two possible interpretations in our model. Either the universe is in its true vacuum and the accelerated expansion is being driven by some quintessence field, the dynamics of an as-yet-undetected ultralight sector, or the universe is in a false vacuum with (fine-tuned) energy very close to the true vacuum. Even in the latter case, our model has the merit of changing the cosmological constant problem from the twofold question of why we are in a false vacuum with very low energy {\it and} why the true vacuum has vanishing energy to the single fine-tuning problem of a false vacuum with energy very close to that dictated by the generic cosmological constant $\Lambda$ in the action.
Note that all of these statements require that the classical dynamics of our model, in particular the constraint equation \eqref{eq:constraint:dual}, can be solved exactly, a requirement that engenders subtleties that we will discuss in \Sec{sec:quantization}.

\section{Implications and Discussion}\label{sec:implications}

We now address some of the subtleties of our model, including locality, quantization, and the dimensionality of spacetime. We further consider its implications for the evolution of the universe and its relation to unimodular gravity.

\subsection{Weinberg's no-go theorem}

An important hurdle to be faced by any model addressing the cosmological constant problem is Weinberg's celebrated no-go theorem \cite{Weinberg:1988cp}. Briefly, the no-go theorem forbids the existence of any solution to the cosmological constant problem within local quantum field theory that does not contain tuning. Let us briefly sketch the proof of the Weinberg no-go theorem as well as observe how the model presented in this paper evades the theorem.

The proof of the theorem proceeds by first assuming a theory described entirely by an action $S=\int \mathrm{d}^4 x \widehat{\mathcal{L}}$, where $\widehat{\mathcal{L}}$ is a local functional that encodes the dynamics of the metric $g_{\mu\nu}$ and $N$ matter fields $\psi_i$. Assuming the theory solves the cosmological constant problem, one then posits the existence of a translation-invariant field configuration (so that the spacetime is Minkowski and the fields are constant), for which the field equations satisfy
\begin{equation}
\frac{\partial \widehat{\mathcal{L}}}{\partial\psi_i} = 0 \qquad \text{and} \qquad \frac{\partial \widehat{\mathcal{L}}}{\partial g_{\mu\nu}} = 0.\label{eq:nogoeom}
\end{equation}
Weinberg then argues that despite the $N+6$ equations of motion and the same number of unknown field values, one cannot obtain a generic (untuned) solution to \Eq{eq:nogoeom}. In particular, after imposing translation invariance, diffeomorphism symmetry is reduced to $\mathrm{GL}(4)$, under which the Lagrangian must transform as $\widehat{\mathcal{L}}\rightarrow \det M \widehat{\mathcal{L}}$ for $M\in \mathrm{GL}(4)$. With \Eq{eq:nogoeom}, one then finds that the Lagrangian must satisfy
\be 
\widehat{\mathcal{L}} = \sqrt{-g} V_0,
\ee
where $V_0$ is a constant independent of the metric. Setting $V_0$ to zero is not accomplished by any equation of motion and the cosmological constant problem thus cannot be solved, within the hypotheses of the theorem, without explicitly assuming a tuning.

In essence, the Weinberg no-go theorem is the statement that the equations of motion of the metric and local matter fields do not set the zero point of the potential within quantum field theory. However, the model presented in this paper evades the hypotheses of the no-go theorem by including a {\it nonlocal} parameter, $\eta$, which couples to the entire action as in \Eq{eq:etaaction}. The equation of motion for this field, \Eq{eq:condition}, constitutes an additional constraint. Importantly, however, it does not add an additional unknown, since as we have seen the value of $\eta$ itself is arbitrary. The equation of motion for $\eta$ thus provides the constraint necessary to set $V_0$ to zero for translation-invariant solutions.

The inclusion of $\eta$ in the action is different than simply tuning $V_0$ to zero, since $\eta$ does more than merely act on the cosmological constant. Indeed, $\eta$ couples to the entire action in the egalitarian manner of \Eq{eq:etaaction}, which in principle could have induced observable changes to cosmological predictions. However, as we have seen, the presence of $\eta$ proves to be compatible with cosmological observations and in fact yields predictions for the nature of the present epoch of accelerated expansion, as noted in \Sec{sec:averages}. In essence, one should view $\eta$ as a convenient mechanism for enforcing the vanishing of the action, integrated over all of spacetime. If the vanishing of the action is indeed to be a fundamental principle, its origin must come from some deeper mechanism, about which we remain agnostic in this paper.

\subsection{Locality}

The question of locality arises naturally in a model that contains averages of fields over all of spacetime.
Any constraint on the action that does not come from a local field is necessarily nonlocal, in the sense that the constraint is an integral equation over all of spacetime.
Whether this is an issue for causality should in practice be a question of whether one can use this fact to construct a causal paradox (i.e., form a closed signal trajectory or have noncommuting operators outside the lightcone).
As we have shown, however, the theory obeys energy conservation, leaves tests of gravity unmodified, and, for an infinite universe, when spacetime averages over the matter fields vanish, reduces to the standard Einstein equation with zero cosmological constant.
Hence, the model seems to yield no ability to produce a causal paradox or indeed even give an observable test. 
In particular, the only place that acausality/nonlocality is present in our model is in setting the value of the effective cosmological constant equal to $\langle R/2 + 8 \pi G \,\mathcal{L}_\mathrm{m}\rangle$.

While the presence of spacetime averages in equations of motion is unusual, field equations containing spacetime averages, as in \Eq{eq:Einstein}, have been obtained in other contexts, such as \Ref{ArkaniHamed:2002fu} (which proposed addressing the cosmological constant problem by effectively turning Newton's constant into a high-pass filter),  \Ref{Kaloper:2013zca} (which replaced the metric in the matter sector with one rescaled by a Lagrange multiplier akin to our $\eta$), and \Ref{Tseytlin:1990hn} (which replaced the action to be varied over by an effective action that was divided by the spacetime volume).

Interestingly, in a universe with positive curvature, matter, and nothing else, our model predicts an observed negative cosmological constant with magnitude equal to $4\pi G$ times the matter density averaged over the life of the universe, which for universes that recollapse soon enough would be observable; however, our universe, by its observed expansion and vanishing curvature, does not fall into this category.

\subsection{Relation to unimodular gravity}

It is useful to consider the relationship between our model and another paradigm for addressing the cosmological constant problem, namely, unimodular gravity \cite{Einstein:1919gv,vanderBij:1981ym,Buchmuller:1988yn,Henneaux:1989zc,Ellis:2010uc}. The similarity between our model and unimodular gravity lies in the fact that both rely on the idea of treating the volume form differently than the spacetime metric. 

In our model, the volume form is dynamical and is simply required to be a four-form $\boldsymbol{\mu}$ that is covariantly constant with respect to the spacetime metric. As we have seen, the extra freedom in the overall scale of the volume form produces an equation of motion that acts as a constraint on the spacetime average of the Lagrangian. Due to the presence of the four-form gauge field $\mathbf{F}$, this constraint equation can be straightforwardly satisfied without appreciably interfering with the usual behavior of gravity on cosmological and astrophysical scales, as we saw in \Secs{sec:GR}{sec:averages}. In particular, the constraint sets the effective value of the cosmological constant to zero, while still allowing for finite periods of inflation.

In contrast, in unimodular gravity, one requires that the volume form be entirely nondynamical. Thus, in computing the equation of motion by varying the metric by $\delta g_{\mu\nu}$, one must restrict to traceless variations for which $g^{\mu\nu}\delta g_{\mu\nu} = 0$. As a result, the equation of motion one obtains is the trace-subtracted Einstein equation 
\be
R_{\mu\nu} - \frac{1}{4} R g_{\mu\nu} = 8\pi G\,\left(T_{\mu\nu} - \frac{1}{4} g_{\mu\nu} T\right).\label{eq:unimodGR}
\ee
The cosmological constant term in the action, which couples to gravity only through its multiplication by the volume form, thus entirely decouples in unimodular gravity. Furthermore, unlike in general relativity, conservation of energy-momentum $\nabla_\mu T^{\mu\nu} = 0$ does not follow automatically from the Bianchi identity  $\nabla^\mu R_{\mu\nu} = \nabla_\nu R/2$, but rather is imposed independently; taking the derivative of both sides of \Eq{eq:unimodGR}, we have
$\nabla_\mu R + 8\pi G \nabla_\mu T = 0 $, so one can identify $R + 8\pi G\,T$ as a constant of integration that can be suggestively labeled $4\Lambda_{\rm int}$. Then \Eq{eq:unimodGR} reduces to the Einstein equation, 
\be 
R_{\mu\nu}-\frac{1}{2}R\,g_{\mu\nu} + \Lambda_{\rm int} g_{\mu\nu}= 8\pi G\,T_{\mu\nu},\label{eq:lambdaint}
\ee
but where, instead of containing the bare cosmological constant in the action, it simply contains a constant of integration. One could then apply the typical anthropic or Euclidean path-integral arguments for why $\Lambda_{\rm int}$ should take a particular value. 

In contrast, our model selects an effective zero cosmological constant via the dynamics of the volume form itself, so the cosmological constant as an arbitrary constant of integration does not arise. Interestingly, taking the trace of the field equation for our model \eqref{eq:Einstein} and then taking the spacetime average, we have
\be
\langle\mathcal{L}_\mathrm{m}\rangle=\frac{1}{4}\left(\langle T \rangle - \frac{1}{8\pi G}\langle R \rangle\right),
\ee
so plugging this back into \Eq{eq:Einstein}, we have
\be
R_{\mu\nu} - \frac{1}{2} R g_{\mu\nu} + \frac{1}{4} \langle R \rangle g_{\mu\nu} = 8\pi G \left( T_{\mu\nu} - \frac{1}{4}\langle T \rangle g_{\mu\nu}\right).\label{eq:prelim}
\ee
Taking the trace once more, we find that
\be 
R + 8\pi G\,T = \langle R + 8\pi G \, T \rangle,\label{eq:prelim2}
\ee 
so we are able to eliminate the spacetime averages from \Eq{eq:prelim}, recovering the equation
\be
R_{\mu\nu} - \frac{1}{4} R g_{\mu\nu} = 8\pi G \left(T_{\mu\nu} - \frac{1}{4} T g_{\mu\nu}\right). 
\ee
That is, our model predicts the trace-free Einstein equation \eqref{eq:unimodGR} and is thus just as consistent with observations as unimodular gravity. However, rather than finding that the cosmological constant is a constant of integration as in \Eq{eq:lambdaint}, our model predicts that it is constrained by the dynamics to equal $\langle R + 8\pi G\, T \rangle/4 = \langle R/2 + 8 \pi G\,\mathcal{L}_\mathrm{m}\rangle$. That is, in addition to the trace-free Einstein equation, our model yields the additional equations of motion given by Eqs.~\eqref{eq:constraint:dual}, \eqref{eq:prelim}, and \eqref{eq:prelim2}.

\subsection{Four-form quantization}\label{sec:quantization}

The preceding discussion has been entirely about the classical theory. We now turn to the question of quantization of the four-form, as discussed in \Ref{BoussoPolchinski}, and its relation to the dynamics of our model. In a spacetime of purely four dimensions (i.e., four-dimensional on all length scales), the four-form $\mathbf{F}=\mathbf{F}_4$ is nondynamical to arbitrarily short wavelengths and its dual $-\theta=\mathbf{F}_0 = \star_4 \mathbf{F}_4$ can take on a continuum of values, allowing the constraint equation \eqref{eq:constraint} to be satisfied exactly. Here, we denote the Hodge dual in $D$ dimensions as $\star_D$, which in this section for clarity we define with respect to the conventional volume form $\boldsymbol{\epsilon}$, not $\boldsymbol{\mu}$. A spacetime of exactly $D=4$ is of course compatible with what we know about the world, in which case the predictions of our model regarding the vacuum discussed at the end of \Sec{sec:averages} are in force.

However, if there are more than four dimensions, as for example in string theory, then the story changes somewhat. In eleven spacetime dimensions, one has a four-form $\mathbf{F}_4$ and its dual $\mathbf{F}_7 =- \star_{11} \mathbf{F}_4$. If the spacetime geometry is $K \otimes M_4$, where $M_4$ is a four-dimensional noncompact geometry and $K$ is a spacelike seven-dimensional compact manifold, then the eleven-dimensional four-form can remain a four-form in the effective field theory on $M_4$, while the seven-form becomes a zero-form in the compactified theory. Moreover, string theory naturally produces 5-branes and 2-branes, to which the gauge fields $\mathbf{A}_6$ and $\mathbf{A}_3$ respectively couple, where $\mathbf{F}_n = \mathbf{dA}_{n-1}$. As a result, the usual Dirac quantization condition applies, meaning that if we look at solutions in which the seven-form wraps $K$, we require
\be 
\int_K \mathbf{F}_7 = 2\pi n\text{ for some }n\in \mathbb{Z},
\ee
where in this subsection we absorb the charge into the definition of the gauge field for simplicity. The equation of motion for $\mathbf{F}_7$ implies that $\mathbf{F}_7 = f_7 \boldsymbol{\epsilon}_7$, where $\boldsymbol{\epsilon}_7$ is the seven-dimensional volume form, so $f_7 V_K = 2\pi n$, where $V_K$ is the volume of $K$. The four-form becomes
\be
\mathbf{F}_4 = \star_{11} \mathbf{F}_7 = f_7 \boldsymbol{\epsilon}_4 = \frac{2\pi n}{V_K} \boldsymbol{\epsilon}_4. 
\ee
Thus, the value of the four-form is quantized, in units set by the volume of the compactified space. 

Naively, this quantization condition on the allowed classical values of the four-form is an obstacle to our mechanism, in which the four-form adjusts to cancel the cosmological constant exactly. On closer examination, however, such quantization can be completely compatible with our approach.

In \Ref{BoussoPolchinski} it was shown that by adding multiple four-forms to the theory, there existed solutions in which the effective step size of the quantization---and hence the effective minimal size of the cosmological constant for fixed $V_K$---could be decreased. Let us suppose that in our model we have $\mathcal{O}(10^2)$ different four-form fluxes, which \Ref{BoussoPolchinski} argued is both plausible in string theory and sufficient to cancel the bare cosmological constant to within the value suggested by observations, and let $\Delta_\Lambda$ be the difference, i.e., the minimal effective cosmological constant in the action. 
The equations of motion in our model, in particular the constraint equation \eqref{eq:constraint} coming from the dynamics of $\eta$, arise from the usual assumption that the principle of least action dictates that the classical dynamics must be a saddle point of the action as a functional of the fields. While this is valid classically, it does not take into account the possibility explored in this section that quantum mechanics may dictate a discretuum of field values, in this case for the four-form flux. This means that the principle of least action, as applied to this system, selects field configurations in which the variation of $S$ is nonvanishing, but minimized. 

Typically, the dominant contribution to the path integral comes from trajectories for which the action is stationary, so that the phases for nearby paths do not destructively interfere. However, if quantization of the four-form flux makes it impossible for the equation of motion for $\eta$ in \Eq{eq:constraint} to be exactly satisfied, then these saddle points are forbidden and the path integral will be dominated by field configurations for which nearby paths interfere least destructively, i.e., those for which $\delta S$ is minimized. That is, the equation of motion for $\eta$ requires field strengths for the four-forms that minimize the effective cosmological constant. We then have $\delta S \propto \Delta_\Lambda$ when the fields are varied. Replacing the zero on the right side of \Eq{eq:constraint:dual} with $\mathcal{O}(\Delta_\Lambda)$ adds a cosmological constant of order $\Delta_\Lambda$ to the field equation~\eqref{eq:Einstein}. For a positive cosmological constant, the matter Lagrangian and energy-momentum vanish as the universe expands, so taking the spacetime average of \Eq{eq:Einstein} leads to $\langle R \rangle \sim \Delta_\Lambda$ and thus de~Sitter space is a consistent solution. 

\subsection{Generalization to top forms in $D$ dimensions}

We can generalize our model to a spacetime $\mathcal{M}$ of arbitrary dimension $D>4$, which will eventually be compactified on a manifold $K$ of $D-4$ dimensions. Consider a model with a top form $\mathbf{F}_{D}=\mathbf{dA}_{D-1}$ coupled to gravity, with bare $D$-dimensional cosmological constant $\Lambda$ and an overall Lagrange multiplier field $\eta$,
\be
S=\eta\int\mathrm{d}^{D}x\sqrt{-g}\frac{1}{16\pi G}\left(R-2\Lambda\right)+\frac{1}{2}\eta\int_{\mathcal{M}}\mathbf{F}_{D}\wedge\star_D\mathbf{F}_{D}-\eta\int_{\partial\mathcal{M}}\mathbf{A}_{D-1}\wedge\star_D\mathbf{F}_{D},
\ee
where $\star_D$ is defined here with respect to the volume form associated with the metric, $\boldsymbol{\epsilon}$. The equation of motion for $\mathbf{F}_{D}$ is $\nabla_{a_1}F^{a_{1}\cdots a_{D}}=0$, where we use Latin indices for the $D$-dimensional spacetime. Hence, $F_{a_{1}\cdots a_{D}}=\theta\epsilon_{a_{1}\cdots a_{D}}$ for $\theta=\text{constant}$. We have $F_{a_{1}\cdots a_{D}}=D\nabla_{[a_{1}}A_{a_{2}\cdots a_{D}]}$.

The equation of motion for $\eta$ sets the spacetime average of the Lagrangian density to zero. The canonical normalization of the $D$-form in terms of its components is
\be 
\mathcal{L}_{F}+\mathcal{L}_{DJ}=-\frac{1}{2\cdot D!}F_{a_{1}\cdots a_{D}}F^{a_{1}\cdots a_{D}}+\frac{1}{\left(D-1\right)!}\nabla_{a_{1}}\left(F^{a_{1}\cdots a_{D}}A_{a_{2}\cdots a_{D}}\right),
\ee
so dualizing, we have the on-shell Lagrangians just as in \Eq{eq:onshellL},
\be 
\begin{aligned}
\mathcal{L}_{F}&=\frac{1}{2}\theta^{2}\\
\mathcal{L}_{DJ}&=\frac{1}{\left(D-1\right)!}\nabla_{a_{1}}\left(F^{a_{1}\cdots a_{D}}\right)A_{a_{2}\cdots a_{D}}+\frac{1}{D!}F_{a_{1}\cdots a_{D}}F^{a_{1}\cdots a_{D}}=-\theta^{2},
\end{aligned}
\ee
where we used the fact that $\epsilon_{a_{1}\cdots a_{D}}\epsilon^{a_{1}\cdots a_{D}}=-D!$ in mostly-plus signature. Thus, the constraint from $\eta$ becomes
\be 
\frac{1}{16\pi G}\left(\left\langle R\right\rangle -2\Lambda\right)-\frac{1}{2}\theta^{2}=0.\label{eq:Dconstraint}
\ee
Meanwhile, the equation of motion for $g_{ab}$ is
\be 
\frac{1}{16\pi G}\left(R_{ab}-\frac{1}{2}Rg_{ab}+\Lambda g_{ab}\right)+\frac{1}{4\cdot D!}g_{ab}F_{a_{1}\cdots a_{D}}F^{a_{1}\cdots a_{D}}-\frac{1}{2\left(D-1\right)!}F_{aa_{2}\cdots a_{D}}F_{b}^{\;\;a_{2}\cdots a_{D}}=0,
\ee
so dualizing using the identity $\epsilon_{aa_{2}\cdots a_{D}}\epsilon^{ba_{2}\cdots a_{D}}=-(D-1)!\delta_{a}^{b}$, we have
\be 
\frac{1}{16\pi G}\left(R_{ab}-\frac{1}{2}Rg_{ab}+\Lambda g_{ab}\right)+\frac{1}{4}g_{ab}\theta^{2}=0.\label{eq:D_EFE}
\ee
Hence, putting together \Eq{eq:D_EFE} with \Eq{eq:Dconstraint}, we end up with the same equation of motion as in \Eq{eq:Einstein}, which for vacuum solutions is
\be 
R_{ab}-\frac{1}{2}Rg_{ab}+\frac{1}{2}\left\langle R\right\rangle g_{ab}=0.\label{eq:D_EFEfinal}
\ee
Adding matter would proceed in exactly the same way as in \Sec{sec:model}.

Let us now compactify the spacetime on $K$. We define $V_{K}=\int_{K}\mathrm{d}^{D-4}x\,\sqrt{-g}$. We assume a product metric for simplicity and ignore the degrees of freedom associated with the components that mix compact and noncompact directions, so that $R_{ab}$ is block-diagonal and thus we can define $R=\,^{(4)}\!R+R_{K}$, where $R_{K}$ goes as the contraction of the components of the Ricci tensor along $K$. The low-energy gravitational sector action becomes
\be 
S_{G}=\frac{V_{K}}{16\pi G}\int\mathrm{d}^{4}x\sqrt{-^{(4)}g}\left(\left\langle R_{K}\right\rangle +^{(4)}\!R-2\Lambda\right),
\ee
where $\left\langle R_{K}\right\rangle$ is the value of $R_{K}$ averaged over $K$, which is equivalent to its value averaged over all spacetime, since $R_{K}$ is by definition independent of the noncompact directions. The on-shell value of the Lagrangian for the $D$-forms in the four-dimensional effective theory is $V_{K} \theta^{2}/2$. That is, everything except the matter action is rescaled by $V_{K}$, so the cancellation of the bare cosmological constant works the same as in the purely four-dimensional model of \Sec{sec:model}. However, there is now an additional term to the cosmological constant, namely, $\langle R_{K} \rangle$, the average of the Ricci scalar on the compact space, which to the four-dimensional effective theory appears as a constant. This additional contribution is canceled along with the bare constant $\Lambda$ by the equation of motion for $\eta$.

We can arrive at this result in another way, by considering the $D$-dimensional Einstein equation \eqref{eq:D_EFEfinal}, evaluated on the noncompact four-dimensional subspace (which we will parameterize with Greek indices),
\be
^{(4)}R_{\mu\nu}-\frac{1}{2}\left(^{(4)}R+\left\langle R_{K}\right\rangle \right)g_{\mu\nu}+\frac{1}{2}\left\langle ^{(4)}R+\left\langle R_{K}\right\rangle \right\rangle g_{\mu\nu}=0,
\ee
where we consider vacuum solutions for simplicity. (If we added matter, the coupling would go as $G_4=G/V_{K}$.) That is,
\be 
^{(4)}R_{\mu\nu}-\frac{1}{2}{}^{(4)}R\,g_{\mu\nu}+\frac{1}{2}\left\langle ^{(4)}R\right\rangle g_{\mu\nu}=0.
\ee
Now, the value of $\theta$ is not quantized, unlike the situation in \Sec{sec:quantization}, since the top form $\mathbf{F}_{D}$ can take on arbitrary values (as its dual, a zero-form, does not wrap any manifold).

\section{Conclusions}\label{sec:conclusion}

We have presented a simple model in which a Lagrange multiplier parameter $\eta$ allows the energy density of a four-form gauge field strength to exactly cancel the cosmological constant.
The model is unusual in that $\eta$ is not a dynamical field and therefore its effects are nonlocal.
In particular, the equations of motion involve averages over all of spacetime.
Perhaps most obviously, there is no especially good motivation for including such a Lagrange multiplier, other than that it gives us the answer we want.

On the other hand, the model has a number of attractive features.
It cancels the low-energy cosmological constant exactly, at least up to possible corrections due to quantization of the four-form field strength. 
It does so while remaining compatible with Weinberg's no-go theorem and in a way that seems compatible with all known observations.
It has the consequence that the late-time vacuum energy is exactly zero, so that our current period of acceleration is necessarily temporary.
We therefore think the model is worth considering on purely phenomenological grounds.

Models that adjust the cosmological constant to zero are often viewed with suspicion, with some justification. 
Even if such a mechanism sets the vacuum energy to zero classically, it is generally difficult to protect such a value against radiative corrections. The situation here is somewhat unusual in that respect. The heavy lifting in our model is done by the Lagrange multiplier $\eta$, which is nondynamical. We therefore do not expect its dynamics to be subject to loop corrections.

Perhaps a more pressing question is that of the naturalness of the form we chose for how $\eta$ enters the action in \Eq{eq:etaaction}.
We motivated that choice by deriving it from the requirement that the volume form $\boldsymbol{\mu}$ be covariantly constant, which is admittedly ad hoc.
One could certainly imagine, for example, more general actions of the form $f(\eta)\int\mathrm{d}^4 x \sqrt{-g} \mathcal{L}$, in which case the constraint that comes from varying with respect to $\eta$ could be satisfied either by setting the conventional action to zero (as in the version we've been considering) or by setting $\mathrm{d}f/\mathrm{d}\eta=0$. The latter choice is simply equivalent to having no constraint in the first place.
Alternatively, one could imagine different functions of $\eta$ multiplying different terms in the Lagrangian:
\be
  S = \int \mathrm{d}^4 x \sqrt{-g} \left[\sum_i f_i(\eta)\mathcal{L}_i\right] .
\ee
Then the constraint from varying $\eta$ would generically not match the condition that the vacuum energy vanish and we would still have a cosmological constant problem.

There is therefore undoubtedly a choice that we made while constructing the model: that the $\eta$ constraint enforce the vanishing of an otherwise conventional action (with a four-form gauge field). 
In our view, this is best understood as a phenomenological approach to a true dynamical mechanism that is yet to be understood, rather than as a complete theory in its own right.
For example, given that the Feynman path integral sums over terms of the form $\exp(-iS/\hbar)$, perhaps our action-minimization procedure could be derived from a principle that treated Planck's constant $\hbar$ as a Lagrange multiplier.
We leave exploration of this and other possible underlying principles for future work.

\begin{center} 
 {\bf Acknowledgments}
 \end{center}

 \noindent 
We thank Cliff Cheung and John Schwarz for helpful comments, as well as Monica Guica and Eugene Lim for discussions and collaboration at an earlier stage of this work. This research is funded in part by the Walter Burke Institute for Theoretical Physics at Caltech, by DOE grant DE-SC0011632, and by the Foundational Questions Institute. G.N.R. is supported by a Hertz Graduate Fellowship and a NSF Graduate Research Fellowship under Grant No. DGE-1144469.

\bibliographystyle{utphys}
\bibliography{FourForm}
\end{document}